\documentclass[english]{aastex}
\usepackage{hyperref}
\usepackage[T1]{fontenc}
\usepackage[latin9]{inputenc}
\usepackage{textcomp}
\usepackage{amssymb}
\usepackage{graphicx}
\usepackage{floatrow}
\usepackage{enumitem}
\usepackage{multirow}


\def\keywords{\vspace{.5em}
  {\textit{Keywords}:\,\relax%
}}

\usepackage{babel}
\shorttitle{Chemical formation of anionic polyynes}
\shortauthors{Gianturco et al.}

\begin{document}

\title{Formation of anionic $C,N$-bearing chains in the Interstellar medium via reactions of $H^{-}$ with  $HC_xN$ for odd-valued  x from 1 to 7
 }

\author{F. A. Gianturco\altaffilmark{1}, M. Satta\altaffilmark{2}, E. Yurtsever\altaffilmark{3}, R. Wester\altaffilmark{1}}

\altaffiltext{1}{Institut fur Ionenphysik und Angewandte Physik Universitatet Innsbruck, Technikerstr. 25/3, A-6020 Innsbruck, Austria}
\altaffiltext{2}{CNR-ISMN and Dept of Chemistry, The University of Rome Sapienza, P.le A. Moro 5, 00185 Rome, Italy}
\altaffiltext{3}{Dept of Chemistry, Ko\c{c} University, Rumelifeneriyolu, Sariyer, TR-34450, Istanbul,Turkey }

\email{Francesco.Gianturco@uibk.ac.at}

\begin{abstract}
 
  We investigate the relative efficiencies of  low-temperature  chemical reactions in the Interstellar medium (ISM) with $H^-$ anion reacting in the gas phase with cyanopolyyne neutral molecules, leading to the formation of anionic   $C_xN^-$ linear chains of different length and of $H_2$. All the reactions  turn out to be without barriers, highly exothermic reactions which provide a  chemical route to the formation of anionic chains of the same length . Some of the anions have been observed in the dark molecular clouds and  in the diffuse interstellar envelopes.Quantum calculations are carried for the corresponding reactive potential energy surfaces (RPESs) for all the odd-numbered members of the series  (x=1, 3, 5, 7). We employ the  Minimum Energy paths (MEPs) to obtain  the relevant Transition State (TS) configurations and  use the latter within the Variational Transition State ( VTS) model to obtain the chemical  rates. The present results indicate that, at  typical temperatures around 100 K,  a set of significantly larger  rate values exists for x=3 and x=5, while  are  smaller for  $CN^-$ and $C_7N^-$. At those temperatures, however, all the rates turn out to be larger than the  estimates  in the current literature for the Radiative Electron Attachment (REA) rates, thus indicating the greater importance of the present chemical path with respect to   REA processes at those temperatures.  The physical reasons for our findings are discussed in detail and linked with the existing observational findings.
 \end{abstract}


\section{Introduction}
The possibility of having molecular systems formed as stable species with a negative charge (anionic molecules) and to have them exist in detectable amounts in the Interstellar Medium (ISM), more specifically in cold dark clouds environments, was put forward several years ago by various authors  \citep{dalgarno73, sarre80, herbst81, herbst97}). They further suggested that carbon chains and hydrocarbon radicals would have large and positive electron affinities (EA) and could therefore be more likely to lead to anion formations via the energy-releasing mechanism of Radiative Electron Attachment (REA). The latter process would occur from the interactions of such molecules with the  free electrons generated in the diffuse regions  by H, He photoionization.
  The recombination process would then be:

\begin{equation}
M+e^{-}\rightarrow\left(M^{-}\right)^{*}\rightarrow M^{-}+h\nu
\end{equation}

They further surmised  that the above processes would provide  efficient mechanisms  for stabilizing molecular anions when the initial neutrals were made up of more than four or five atoms  \citep{herbst09}). The experimental, astrochemical evidence for the presence of such species was only provided later on, when an unidentified series of lines were detected in a radio astronomical survey of the evolved carbon star IRC+10216 by Kawaguchi et al. \citep{Kawaguchi95}, and was conclusively assigned to the spectrum of $C_{6}H^{-}$ by Mc.Carthy et al. \citep{mccarthy06}) on the basis of laboratory rotational spectroscopy and separate observations toward the Taurus Molecular Clouds 1 (TMC-1(CP)).
Additional anionic molecules were  observed later on by \citep{cernicharo07} ($C_{4}H^{-}$), by \citep{remijan07} ($C_{8}H^{-}$) and by  \citep{thaddeus08} $C_{3}N^{-}$) in the envelope of IRC+10216 . In the TMC-1 the $C_{8}H^{-}$ anions was additionally observed by \citep{brunken07}, the largest anionic  molecular chain observed thus far.  Additional observations on molecular negative ions were  surveyed by \citep{gupta07} who observed $C_{6}H^{-}$ in two further sources: the pre-stellar cloud L1544 and the protostellar object L1521F.

Since polyatomic molecules  with  a negative charge have turned out to be present under different conditions and in significant quantities (e.g., the measured column density of $C_{6}H^{-}$ in the TMC-1(CP) was found by \citep{mccarthy06} to be $10^{11}$ cm$^{-2}$), it is reasonable to expect that they also play a significant role in the chemistry of the dark molecular clouds  and at the low temperatures assigned to these regions. Hence, it becomes relevant to investigate in more  detail additional  possible formation paths of such stable anionic species by chemical routes.

From the theoretical and computational standpoints we have already studied in recent years, although on different molecular partners, the general dynamics which could be driven by the free electrons available in that environment and the important role played by intermediate, metastable anions as  molecular gateways to the final, anionic species which are  experimentally detected .
For example, the resonant attachment of slow electrons to $NCCN$ has been studied a while ago in our group \citep{sebastianelli10}), while we  have also analysed the possible  fragmentation decays of $HC_{3}N$ and $HC_{5}N$ upon electron attachment  \citep{sebastianelli12}), as well as carrying out a detailed study of resonant electron attachment to $HC_{4}H$ \citep{baccarelli13}).
A more extensive analysis on the dynamics of electron attachment to non-polar hydrocarbon chains like $HC_{n}H$ (with n from 4 to 12) at the expected low-energies of planetary atmospheres and dark molecular clouds, was also carried out by us  \citep{carelli13}). In that work  it was specifically shown that the attachment mechanism is driven  by the prior formation of metastable anions from asymmetrically deformed  non-polar polyynes, giving rise to complex intermediates of polar radical anions plus H-atom stretched configurations:

\begin{equation}
HC_{6}H+e^{-}\rightarrow\left(HC_{6}H^{-}\right)^{*}\rightarrow H-(C_{6}H^{-*}) \rightarrow C_{6}H^{-} + H 
\end{equation}

where the intermediate system  can initially form its closed-shell anion as  a dipole-bound state (DBS) partner. The latter can  then decay into the more stable anionic valence-bound states (VBS), after internal-energy redistribution  and  ejection of one terminal hydrogen atom.
      The above chain of events is reminiscent of the indirect REA mechanism (IREA) introduced more recently by \citep{douguet13}, where  the dissipation paths of the large amount of energy which has to be released upon electron attachment in systems with large and positive EA values was suggested to occur via an electron-vibration nonadiabatic coupling mechanism. They have shown however that this IREA mechanism, and indeed also the REA mechanism of eq. (1), is unfortunately rather inefficient for the smaller radical chains like $C_{2}H^{-}$ and $C_{4}H^{-}$, thereby calling into question the relevance of such pathways for the formation of observable quantities of the anions of the present molecules. It is therefore  important to further explore and understand which other mechanisms could be  responsible for the formation of the smaller $(C,N)$-chains anions under ISM conditions.

To extend the chemical formation option, we wish to  explore in the present study a  "chemical"  stabilization of the  anions which involve four of the members of the $C$-bearing linear chains that terminate with the cyano group \citep{marisawa05}).
 We shall therefore look  at the efficiency of the formation of such  anions by examining the reaction of $H^{-}$ with   non-polar precursors of the cyanopolyyne series, the  ($HC_{x}N$) molecules,   in the gas-phase :

\begin{equation}
H^{-}+HC_{x}N\rightarrow H_{2}+C_{x}N^{-}
\end{equation}

where the length of the odd-numbered chains goes in our study  from $x=1$  to $x=7$ .

The importance of $H^{-}$ as a chemical partner has been suggested before, given its likely existence through cosmic rays which act during ion-pair reactions within the inner cores (radius $\leq$ 10$^{6}$ $AU$) of the prestellar envelopes  \citep{prasad80}), although little is known so far on the details of its actual mechanism under the above conditions  \citep{mackay77}).

Therefore, if the presence of $H^{-}$ could be  in sufficient amounts to provide a useful chemical partner, then we shall show that  the above reaction is indeed an interesting alternative. They are in fact, as discussed below, markedly exothermic processes where the final product formations involve an electron transfer (ET) mechanism accompanied by the formation of a new $H_{2}$  bond into a separate molecular product from the final anion.

In an earlier study  on the smaller members of the cyano derivatives, $CN^{-}$ and $C_{3}N^{-}$  \citep{satta15} we have  indicated that the chemical nature of their reactions with $H^{-}$, i.e. the suggested  mechanism of  an ET process occurring along with the energy release from the $H_{2}$ formation,  makes the production of their anions to be exothermic and  to occur without an intermediate energy  barrier between reactants and products, although a configurational Transition State (TS) is observed. 
  It thus becomes  interesting to computationally extend the study of these reactions  to verify  their behaviour for larger members of the series, as well as revisiting the earlier results by using an improved description of the quantum chemical interaction forces. 

In the following sections we shall investigate this point for  the reactions involving the first four members of the cyanopolyyne series. The next Section 3 analyses the overall shapes of the various RPESs, while Section 4 calculates the corresponding reaction rates over a  range of temperatures representing   the colder dark cloud conditions and the warmer circumstellar envelopes which are of  interest in our  study. 
 Section 5 will give our present conclusions and also link them with  the existing observational studies on the  abundances of the cyanopolyyne anions.

\section{ Quantum chemical calculations}

Before starting our analysis of the dynamical features of the reactive partners in eq. (3) we need to look  into the quantum structural properties that we have obtained for all partners in the considered processes. All the ab initio structural  calculations were carried out using the MOLPRO suite of codes \citep{werner06}), unless otherwise stated.  In order to better analyse visible trends along the present series of linear anions, we have repeated the calculations for the smaller terms of the series, the $x=1$ and the $x=3$ molecular anions already investigated before \citep{satta15}). We have also extended the analysis of the kinetics of $CN^{-}$ formation by considering both isomeric structures of the initial molecule,$HCN$ and $HNC$. More details on this aspect of the discussion will be given below.

The method employed here has been the Coupled-Cluster-Single-Double-Perturbative-Triple-Excitation ( CCSD(T)) and the quality of the chosen basis sets is defined as augmented-correlation-consistent, corrected up to triple-zeta expansions : aug-cc-p-VTZ . All the present acronyms are explained in detail in the MOLPRO suite of codes \citep{werner06}). All the molecular reagents, as  already known, turned out to be linear in the most stable ground electronic states which correspond to closed-shell structures. Their optimized geometries are presented in Table \ref{tab1}. We also report in that Table the geometries of the isomer radicals $C_{x}NH$ so that they can be compared with the corresponding structures of the $HC_{x}N$ species.

\begin{table}[H]
\caption{Computed,optimized geometries for the neutral radicals and negative \textsuperscript{1}$\Sigma $ molecules of present study. All species are linear and the distances are in {\AA}}
\begin{tabular}{c|c|c|c|c}
$C_xN^-$&x=1&x=3&x=5&x=7\tabularnewline
\hline
C1C2&---&1.263&1.271&1.275\tabularnewline
C2C3&---&1.364&1.342&1.333\tabularnewline
C3C4&---&---&1.246&1.255\tabularnewline
C4C5&---&---&1.352&1.329\tabularnewline
C5C6&---&---&---&1.247\tabularnewline
C6C7&---&---&---&1.351\tabularnewline
C(last)N&1.191&1.187&1.188&1.188\tabularnewline
\hline
\hline
$HC_xN$&x=1&x=3&x=5&x=7\tabularnewline
\hline
C1C2&---&1.217&1.222&1.224\tabularnewline
C2C3&---&1.372&1.359&1.354\tabularnewline
C3C4&---&---&1.228&1.235\tabularnewline
C4C5&---&---&1.364&1.347\tabularnewline
C5C6&---&---&---&1.233\tabularnewline
C6C7&---&---&---&1.361\tabularnewline
C(last)N&1.167&1.177&1.180&1.181\tabularnewline
C1H&1.065&1.063&1.064&1.063\tabularnewline
\hline
\hline
$C_xNH$&x=1&x=3&x=5&x=7\tabularnewline
\hline
C1C2&---&1.274&1.284&1.289\tabularnewline
C2C3&---&1.313&1.305&1.303\tabularnewline
C3C4&---&---&1.263&1.271\tabularnewline
C4C5&---&---&1.300&1.292\tabularnewline
C5C6&---&---&---&1.269\tabularnewline
C6C7&---&---&---&1.298\tabularnewline
C(last)N&1.177&1.178&1.184&1.187\tabularnewline
NH&0.998&0.995&0.996&0.996\tabularnewline
\end{tabular}
\label{tab1}
\end{table}

One clearly sees from the Table's  data that only minor differences in the geometries occur when going from the $HC_xN$ to $C_xHN$: we shall further analyse below the possible effects on the energy changes of these exoergic reactions. The total energies for the ground states of reagents and products are reported in Table \ref{tab2}

\begin{table}[H]
\caption{Computed ground state total energies for reactants, using the CCSD(T) method described in the main text. All values are in units of hartrees.}
\begin{tabular}{c|c}
Molecule&aug-cc-pVTZ energy\\
$H_2$&-1.172613\\
$H^-$&-0.526562\\
\hline
$C_3N^-$&-168.723503\\
$C_5N^-$&-244.743867\\
$C_7N^-$&-320.762897\\
\hline
$HC_3N$&-169.288765\\
$HC_7N$&-245.302425\\
$HC_7N$&-321.317307\\
\end{tabular}
\label{tab2}
\end{table}

As mentioned earlier, all the present reactions are exothermic processes involving polar targets. To check on the quality of our description of them, we report in Table \ref{tab3} the calculated values of the permanent dipoles of the molecular partners undergoing the electron transfer (ET) process during the reactions.

\begin{table}[H]
\caption{Computed permanent dipole moments (in units of Debye) for the neutral counterparts of the final product. RODFT, ROMP2 and ROCCSD(T) denote restricted-open-shell calculations.  $^*$Anion values are from \cite{care14}}

\begin{tabular}{|c|c|c|c|c|}
\hline 
 CN& RODFT(B3LYP) & ROMP2 & ROCCSD(T) & Anion{*}\tabularnewline
\hline 
\hline 
6-311G(2df,2pd) & -1.3058 & -2.2765 & -2.2569 & 0.69\tabularnewline
\hline 
6-311++G(3df,3pd) & -1.3811 & -2.3256 & -2.3087 & \tabularnewline
\hline 
cc-pVTZ & -1.3222 & -2.2847 & -2.2672 & \tabularnewline
\hline 
aug-cc-pVTZ & -1.3794 & -2.3233 & -2.3069 & \tabularnewline
\hline 
\end{tabular}

\begin{tabular}{|c|c|c|c|c|}
  \hline
  \hline
$C_3N$ & ---- &  ----- &  -----      & Anion{*}\tabularnewline
\hline 
\hline 
6-311G(2df,2pd) & -2.8394 & -3.1759 & -3.1753 & 2.72\tabularnewline
\hline 
6-311++G(3df,3pd) & -2.9571 & -3.2539 & -3.2533 & \tabularnewline
\hline 
cc-pVTZ & -2.8807 & -3.2058 & -3.2052 & \tabularnewline
\hline 
aug-cc-pVTZ & -2.9552 & -3.3252 & -3.2506 & \tabularnewline
\hline 
cc-pVQZ & -2.9340 &  &  & \tabularnewline
\hline 
aug-cc-pVQZ & -2.9554 &  &  & \tabularnewline
\hline 
\end{tabular}

\begin{tabular}{|c|c|c|c|c|}
  \hline
  \hline 
$C_5N$ &   ---   &  ----   &    ----      & Anion{*}\tabularnewline
\hline 
\hline 
6-311G(2df,2pd) & 1.2490 & -0.3117 &  & 5.61\tabularnewline
\hline 
6-311++G(3df,3pd) & 1.2796 & -0.3310 &  & \tabularnewline
\hline 
cc-pVTZ & 1.2997 & -0.2992 &  & \tabularnewline
\hline 
aug-cc-pVTZ & 1.2932 & -0.3181 &  & \tabularnewline
\hline 
\end{tabular}

\begin{tabular}{|c|c|c|c|c|}
  \hline
  \hline 
$C_7N$ & ----  & ---- & ---- & Anion{*}\tabularnewline
\hline 
\hline 
6-311G(2df,2pd) & 0.7796 & -0.4389 & -0.4389 & 4.19\tabularnewline
\hline 
6-311++G(3df,3pd) & 0.7888 & -0.4648 &  & \tabularnewline
\hline 
cc-pVTZ & 0.8083 & -0.4352 & -0.4352 & \tabularnewline
\hline 
aug-cc-pVTZ & 0.7978 & -0.4541 &  & \tabularnewline
\hline 
\end{tabular}

\label{tab3}
\end{table}

It is interesting to note that the dipole moment of the CN radical is found, ensuing the largest basis set in the bottom line in the Table, to vary between -1.379 and the CCSD(T) approach (last column to the right) of -2.307 Debye. The experimental value of a while ago is 1.45 \cite{thom68} while a recent calculation \cite{fart11} yielded a value of -1.47.

For the $C_3N$, $C_5N$ and $C_7N$ the largest basis set was used within the Density Functional approach (DFT) using the B3LYP exchange functional model, providing decreasing values of -2.95, +0.80 and +1.30 Debye in Table \ref{tab3}. Our earlier calculations had yielded values of -3.0, +0.80 and 1.29: no experimental values or earlier calculations  exist on the radicals, as discussed by  \cite{care14}. On the other hand, all the corresponding anions have very different dipole moment values, as reported in the last columns of table \ref{tab3} and as computed earlier by  \cite{care14}: no experimental values are available for the closed-shell anions, while earlier calculations by \cite{bots08} and by  \cite{kolo08} yielded dipole moment values with the same orientation (i.e. along the z-axis and now pointing from the N to the C atoms in each member of the series) and about 20\% larger.

In all cases, therefore, the dipole moment values increase dramatically in going from the radical to the closed-shell anionic structures. Along the same series of cyanopolyynes the corresponding electron-affinities (EA) also increase rather markedly from the n=1 member (+3.8 eV) to the n=7 member (+4.61 eV) as discussed already in our earlier study \cite{care14}. This means that the occurrence of the  ET step during the reactions we are going to discuss affords large energy releases when the electron moves from the $H^-$ partner (with an EA value of 0.754 eV \cite{shie00}) to the N-end of the final linear chains with much larger EA values. This aspect will be further discussed below in more detail.

We have computed again the EA values at the equilibrium geometries of the present radicals and found the results reported below  in Table \ref{tab4}.

\begin{table}[H]
\caption{Computed EA values at the CCSD(T) level of calculation of the present study (eV)}
\begin{tabular}{|c|c|c|c|}
\hline 
 & Present Calculations & Expt(\cite{care14}) & Calculations(\cite{care14})\tabularnewline
\hline 
\hline 
$CN$ & 4.00 & 3.86 & 3.80\tabularnewline
\hline 
$C_3N$ & 4.64 & 4.59 & 4.40\tabularnewline
\hline 
$C_5N$ & 5.09 &  & 4.59\tabularnewline
\hline 
$C_7N$ & 5.51 &  & 4.61\tabularnewline
\hline 
\end{tabular}

\label{tab4}
\end{table}
We can see from the Table that all the present radicals have very large EA values, thus making the ET process from the $H^-$ reagent into the final anions a  strongly exothermic process leading to the TS formation complex before the additional energy release due to the formation of the neutral $H_{2}$  bond during formation of the final molecular products. We shall further illustrate below in more detail the energetics of the present reactions.

If we look at the data at the CCSD(T) level of calculations, we see that all the reactions involving the $HC_xN$ reagents are consistently exothermic along the series, varying by more than 10\% as x increases from 1 to 7: as we shall further see below, none of them shows any barrier from reagents to products, thus confirming their feasibility at the low temperatures of the DMC environments.

Additionally, we see that the same set of reactions involving their isotopologue variants, occurs with the $HNC_x$ series of reagents (lower panel in Table \ref{tab5}).

\begin{table}[H]
\caption{Computed exothermicity values for the two series of reactions involving both isotopologue forms.Units are in eV}

\begin{tabular}{|c|c|c|c|}
\hline 
\multicolumn{4}{|c|}{$H^{-}+HC_{x}N\rightarrow H_{2}+C_{x}N^{-}$}\tabularnewline
\hline 
\hline 
x & MP2 & MP2+ZPE & CCSD(T)\tabularnewline
\hline 
1 & 2.11 & 2.14 & 2.12\tabularnewline
\hline 
3 & 2.17 & 2.19 & 2.20\tabularnewline
\hline 
5 & 2.37 & 2.40 & 2.38\tabularnewline
\hline 
7 & 2.53 & 2.58 & 2.49\tabularnewline
\hline 
\end{tabular}

\begin{tabular}{|c|c|c|c|}
\hline 
\multicolumn{4}{|c|}{$H^{-}+HNC_{x}\rightarrow H_{2}+NC_{x}^{-}$}\tabularnewline
\hline 
\hline 
x & MP2 & MP2+ZPE & CCSD(T)\tabularnewline
\hline 
1 & 2.89 & 2.90 & 2.76\tabularnewline
\hline 
3 & 4.59 & 4.56 & 4.48\tabularnewline
\hline 
5 & 5.19 & 5.17 & 5.14\tabularnewline
\hline 
7 & 5.54 & 5.55 & \tabularnewline
\hline 
\end{tabular}
\label{tab5}
\end{table}

The exothermicity values increase quite markedly: an increase of about  0.64 eV for the HCN/HNC case, to more than 2.60 eV for the $HNC_7$ case. This is an interesting result for which, however, we have little observational information for the members with x from 3 to 7. On the other hand, the HCN/HNC isotopologues have been extensively observed in many astrophysical environments: from diffuse and translucent interstellar clouds in \cite{list01, turn97}, to dense interstellar clouds: see \cite{mily10}, to star-forming regions in \cite{jin15}, to protoplanetary disks: \cite{gran15}, to external galaxies as in  \cite{gao04} as well as in comets: \cite{lis08} and planetary atmospheres: \cite{more11}.

Likewise, the HCN/HNC abundance ratios have also shown marked changes between different astrochemical environments. In the dense interstellar clouds, in fact, the gas is shielded from external UV starlight and the abundance ratio was found to be around 1.0 in \cite{sarr10}. On the other hand, in regions exposed to UV photons HCN was found to be more abundant than HNC by a factor of 5 in both diffuse clouds: \cite{list01} and in PDR environments as in \cite{hoge95}. It therefore stands to reason that both reactions could occur with $H^-$, thereby increasing the probability of forming $CN^-$, as we shall further discuss below.

Although, for the longer members of the series ,we know little about the ISM relative abundances of their isomeric structures , our findings provide already a relevant result in the sense that, within the chemical route that we intend to explore in the present study, our data from Table 5 indicate that both isomers  would  important to investigate in order to consider both  chemical paths to the formationof their corresponding anions. However, we shall mainly discuss in the following the reactions with the $HCN$  and  $HNC$ reagents,  to show these paths to forming  $CN^-$ as being equally possible in the different environments of the ISM  we are analyzing.

\section{The shapes of the reactive surfaces}

\begin{figure}[t!]
\includegraphics [width=\textwidth] {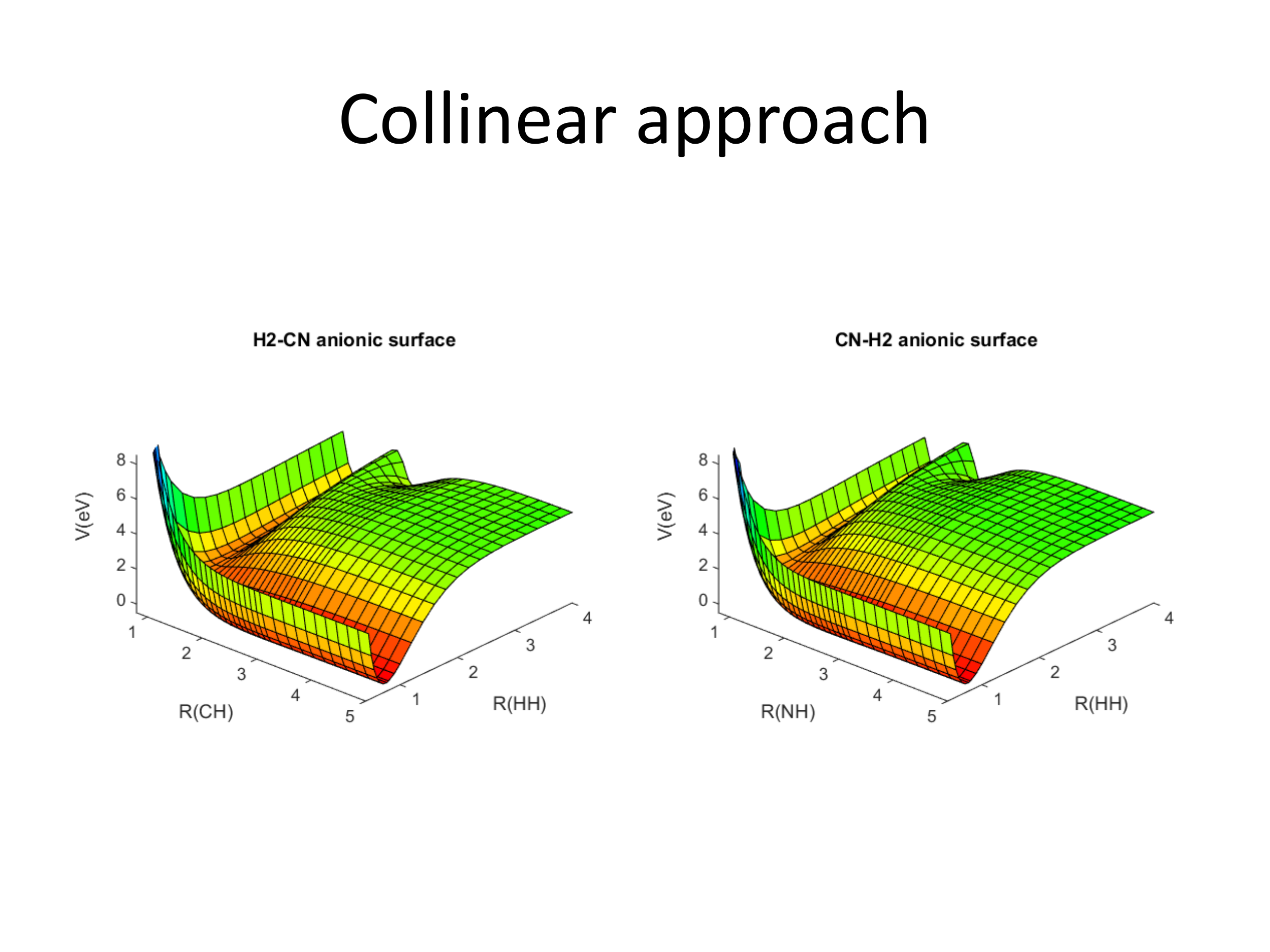}
\caption{Computed Reactive PES for the formation of $CN^{-}$ form either $H^{-}+HCN$ (left panel) or $H^{-}+HNC$ (right panel). See main text for further discussion.}
\label{fig1}
\end{figure}

As discussed before, all reactions of interest in this study are strongly exothermic. Furthermore, our previous experience with such process, as discussed in \cite{gian16, satta15}, has shown us that they are dominated by a collinear Minimum Energy Path (MEP) and  without any barrier existing between reagents and products. The suggested mechanism, in  fact, is given in \cite{gian16, satta15} as one in which the initial  ET process occurs during the approach between the two hydrogen atoms:  the H-C bond is stretched as the H  anion is approaching, thereby  allowing  for the starting of the electron transfer exothermic step from the approaching $H^-$  initially to the C-end of the molecular partner but finally, and chiefly, to the N-end of the radicals ($C_xN^-$).The formation of a TS configuration as the partners approach each other more closely now further  leads  to additional energy release into the final products by forming a neutral , vibrationally cold  $H_2$  molecule that separates from the highly stretched configuration it has within the transition complex.
A pictorial representation of the 3D surface describing the reaction (RPES) is reported by Figure \ref{fig1}, where  the two panels compare the two isomeric variants we have discussed before. The left panel therefore shows the reaction of $H^-$ with HCN, while the one on the right reports the same reaction but with HNC: both sets of reagents yield $H_2$ and $CN^-$ as products.

It is evident from a perusal of both surfaces that the two reactions proceed in the same fashion: no energy barriers between reagents (on the left region of each RPES) and products (on the right region of the same surfaces) and a marked exothermicity which, for the case of the  HNC, is even larger by more than 0.5 eV (see data in table \ref{tab5}). We therefore  expect that both reactions can play a role in the ISM regions where both molecules in question have been observed

It is also  interesting to note here that both energy surfaces show an intermediate, smaller  energy minimum at the stretched distances of the $C-H$ end of the partner molecular chain ( around 2 to 4 {\AA}) and at even larger values of the $H-H$ distances of about 4 {\AA}.  One could suggest that in that outer well region the initial ET step begins to  occur while the molecular hydrogen is  still strongly vibrationally excited and the terminal $H$ atom has moved far away from the residual cyano derivative.. In other words the excess electron would  initially "transfer"  to the H-end of the chain and finally to the CN-end of the new anion after the $H_2$ is formed as one of the molecular products. It is thus at the second energy minimum, which is now deeper and occurs at shorter $C-H/N-H$ distances that the  neutral, $H_2$ molecule starts to get formed  (second exothermic step) by releasing its vibrational energy content and thus reaching the variationally optimized TS which evolves towards finally producing $H_2$ in its vibrational ground state as it  separates  from the anionic product.

\begin{figure}[t!]
\includegraphics [height=12cm] {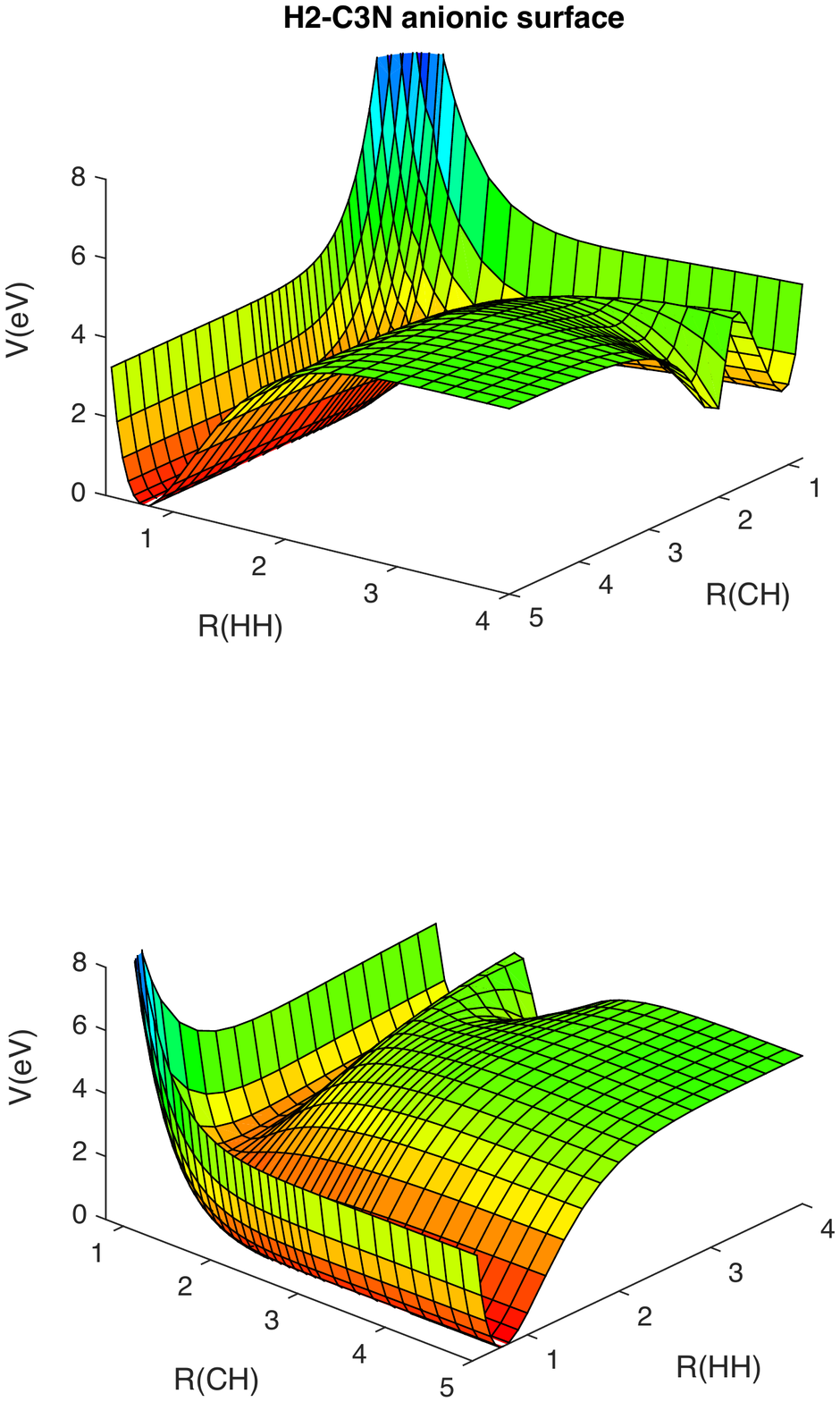}
\caption{Two different views of the reactive PES for the $H^-+HC_3N$ reaction. The upper panel shows the entrance channels on the upper right region of the RPES, while  the lower panel reports the products' channel on the lower right portion of the surface. See main text for further details.}
\label{fig2}
\end{figure}

In the two panels of Fig. \ref{fig2} we further report the RPES shape for the reaction involving the next member of the cyanopolyyne series: the $HC_3N$ neutral partner of $H^-$. The upper panel of the Figure shows in the forefront the $H^-$ approach to the $HC_3N$ reagent (from right to left) and the clear exothermic, barrierless path down to $H_2$ formation and $C_3N^-$ separation on the further valley going out on the left of that panel. On the other hand, the rotated view presented in the lower panel indicates on the extreme left the incoming, exothermic path for the $H^-$ approach to the C-H end of $HC_3N$, followed by the further, exothermic outgoing $C_3N^-$ and of the cold $H_2$ formation along the valley shown in the forefront, from left to right.

\begin{figure}[t!]
\includegraphics [height=12cm] {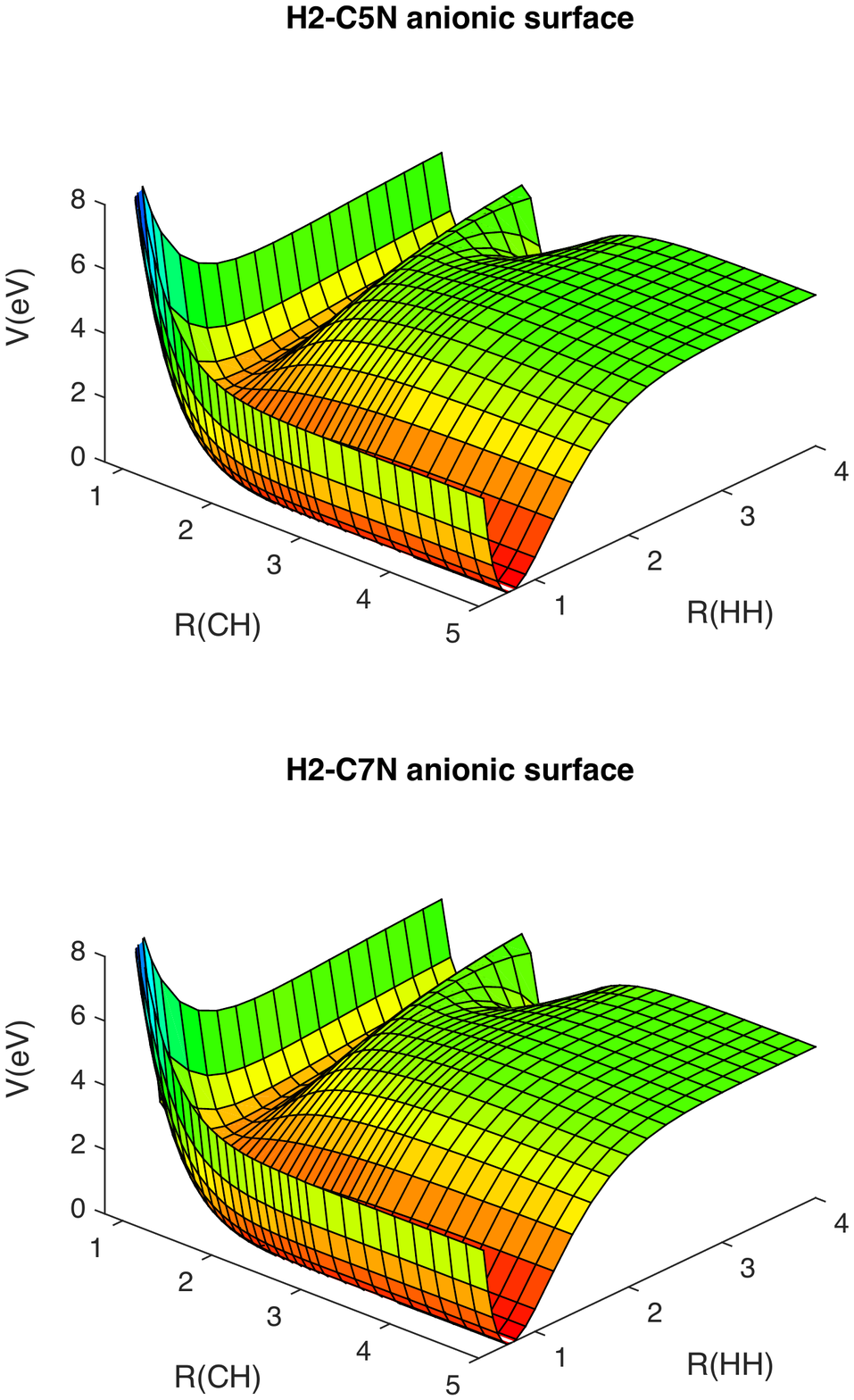}
\caption{Computed 3D presentation for the collinear MEPs relative to the next larger members of the cyanopolyyne series. Upper panel: $H^-+HC_5N$; lower panel:  $H^-+HC_7N$. See main text for further details. }
\label{fig3}
\end{figure}

From the upper view of the RPES shown in that Figure we see even more clearly the role of the outer, shallower energy well that we associate to the initial ET step while the $H-H$ bond is still describing  a highly vibrationally excited pair of H atoms, the external one being also far away from the residual cyano derivative fragment. It is thus an indication of the occurrence of a non-adiabatic curve-crossing effect that transfer the $H^-$ charge to the terminal $H$ atom of the cyanopolyyne partner. . It is then the deeper well located at the shorter C-H distances which will permit  the exothermic release of that vibrational energy along the evolution of the variationally optimized TS complex (as we shall discuss in more detail in the next Section) that can now decay into the final "cold" molecular  products without any intermediate energy barrier and having completed the ET step to the final anionic product. 

In conclusion,  this member of the cyanopolyyne series  also show its reaction with $H^-$ to be a barrierless, exothermic process along a nearly collinear path. Along that path, in fact,  one gains energy by the occurrence of an initial Electron-Tranfer (ET) process which takes place  for highly stretched $C-H$ and  $H-H$ bonds between partners. The release of that vibrational energy in both bonds helps  the TS formation at the bottom of the RPES, at the start of the MEP path.  As we shall further explain in the next Section, along such reaction path the formation of a cold $H-H$ neutral molecule plus the  outgoing  anionic linear chain as one of the products can now occur in a strong exothermic and barrierless fashion.

The same general features of this reaction are also observed along the collinear MEP energy evolution reported by Figure \ref{fig3} for the next  two members of the series: $C_5N^-$ and $C_7N^-$.

 In both instances the process remains, as before, strongly exothermic and without any barrier between reagents and products after the occurrence of the initial ET step indicated by the presence in both systems of the outer, shallower energy well for vibrationally excited $C-H$ $H-H$ distances. Once the deeper well of the RPES is reached at shorter distances within the complex, the systems reach configurations where both the above bonds are shorter and close to their equilibrium values. This means that the reactions can now follow the exothermic MEP evolutions without any intermediate barrier, as we have already explained before.

\begin{figure}[t!]
\includegraphics [height=12cm] {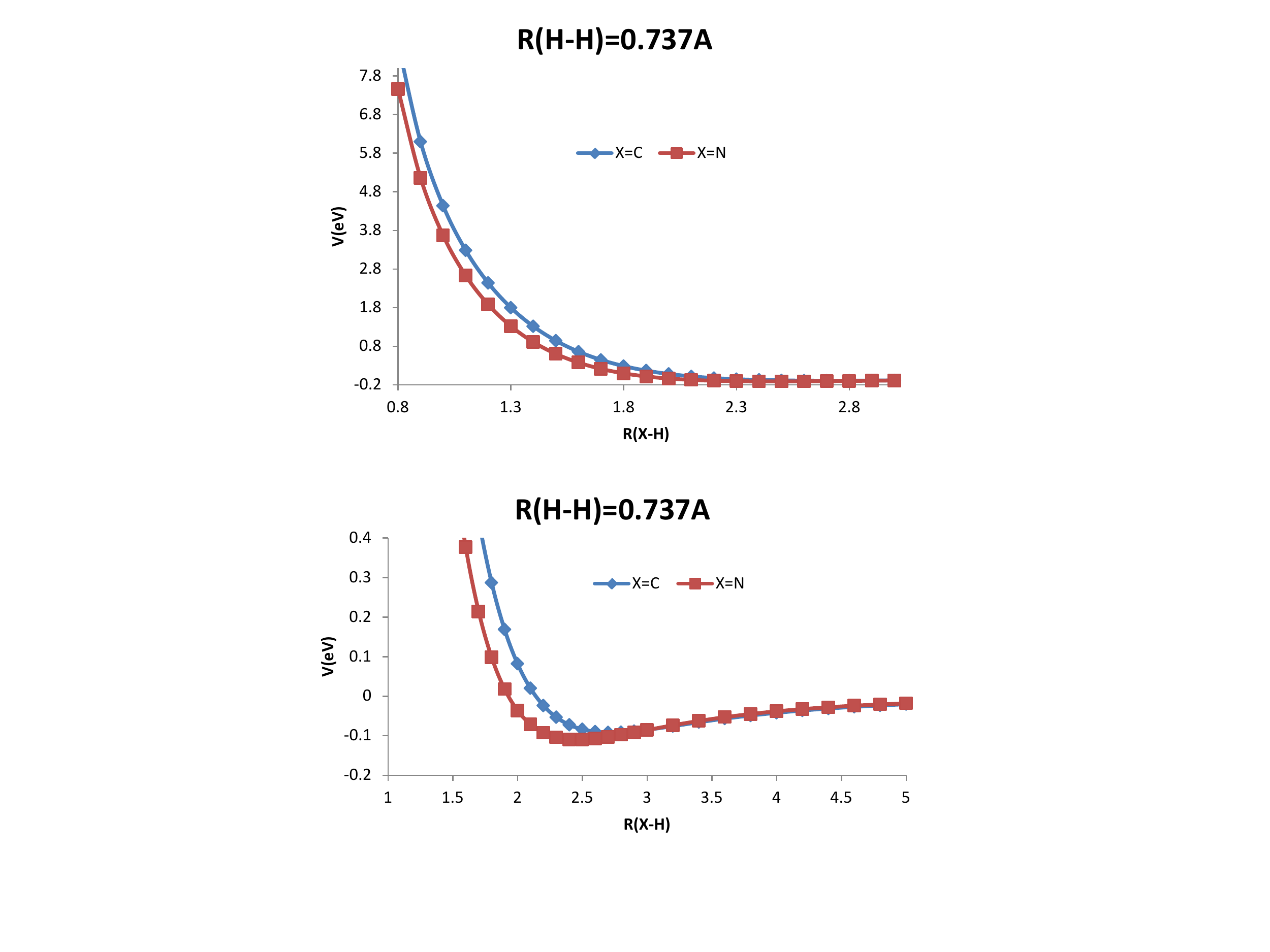}
\caption{Different collinear 'cuts' for the reaction of $H^-+HCN$, given by small lozenges, and $H^-+HNC$ , given by large squares, at fixed H-H distances for the geometry  of the complex.The 'X' symbols denote the remaining chains from either end of approach for the $H^-$ partner.( colours on line).  See main text for further details}
\label{fig4}
\end{figure}

Another interesting way of visualizing the exothermicity of the present reactions is shown by the two panels of Figure \ref{fig4}, where we choose as an example   the isomer reactions involving HCN/HNC partners.

Both panels of that figure present the collinear reaction paths after "cold" $H_2$ formation has occurred and the initial vibrational energy of that bond has been released: the formed anion now recedes from the complex region.  The H-H distance is also  that of the optimized complex  geometry in both cases ( see next Section's discussion). The upper panel clearly shows that in both reactions  the formed anions  leave the complex along a monotonically decreasing path since the formation of a vibrationally cold, neutral $H_2$ now  affords the systems to undergo a substantial energy release without  any intermediate barrier.

We have enlarged in the lower panel of the same figure the energy region where either the H-C or the H-N distances are clearly down into a flat outgoing path. It now shows in both cases the presence of a very shallow well indicating the absolute minimum energy location of the complex configurations before the final release  of the two products, as mentioned already when analysing the previous figures. Although the reaction with HNC shows a slightly larger well (-0.11 eV vs -0.09 eV) into their  configurations of the reaction complex, both systems can easily make out to products even at the lower temperatures, due to the marked shallowness of such small minimum regions. However, this can energetically happen provided the intial ET step has been allowed to occur along the outer wells shown by Figure 2 and already discussed earlier in this Section .

In the next Section we shall be  using  the present RPES  of the colinear approaches to help us to obtain  the corresponding reaction rates over a broad range of temperatures, following the nonlinear, 3D model of the Variational Transition State Theory which we shall  discuss below. in greater detail

\section{Modelling the reaction rates}
We have seen in the previous discussion that all the  reactions evolve on RPESs which are markedly exothermic and present no energy barriers along the MEP evolutions from reagents to products. Furthermore, the calculations of the relative energetics between reagents and products, for all the members of the series we have examined, indicate a marked increase of exothermicity in going from one set of $HC_xN$ partner to the $HNC_x$ isomeric variants. This last feature will be discussed below, as an example, for the case of HCN/HNC systems.

The special features of the reaction energetics have allowed us to use a version of the RRKM approach based on the Variational Transition State Theory (VTST) treatment for obtaining temperature-dependent rates, as discussed in \cite{Fernandez-Ramos06} for the case of exothermic, barrierless reactions. This approach is basically assuming that, for the present types of reactions, the formation of a transition-state complex along the exothermic energy path from reactants to products controls the efficiency of product formation via the relative energetics between the partition functions of that complex and those of the final products.

In particular, within the RRKM approach for the description of a TS configuration, one assumes a strong-coupling approximation whereby the degrees of freedom within the TS complex are strongly coupled among themselves and therefore the entire phase-space available will be occupied by the partition functions (PFs) associated to the TS on a shorter time-scale than that of the characteristic reaction time step: see \cite{Fernandez-Ramos06}.
This means that, once the TS complex  is reached by the reagents along the exothermic MEPs, the reactants are in microcanonical equilibrium. The TS complex partition function, $Q_{TS}^\ddagger$ ,  can then be obtained as the product of the PFs for the conserved mode partition functions within that complex, $Q_{cons}^\ddagger$, times the PFs of the translational modes which are evolving along the MEP following the complex geometry,$Q_{trans}^\ddagger$. The latter modes now make the complex evolve from reagents to products, thereby acting as the active modes of the reaction:

\begin{equation}
Q_{TS}^{\ddagger}(T)=Q_{cons}^{\ddagger}(T)Q_{trans}^{\ddagger}(T)
\end{equation}

In our case the translational modes for the title reactions will be those indicated in the  3D energy  maps of the previous Section: the $H^- \cdots H$ distance and the $H \cdots C$ distance along the collinear path. We shall see later that such simpler approach holds well for the x=1 case, while the larger chains will require a multidimensional optimization along increasingly  non-linear TS structures formed along their respective MEPs.
The pre-exponential factor of the reaction rate formulations becomes now the only significant part  which needs evaluation in order to obtain the required reaction rates $K(T)$. Thus, the latter quantities can now be obtained as the ratios between the relevant PFs of the TS complex and those of the reagents, as a function of the reaction temperatures:

\begin{equation}
  K(T)=\frac{k_BT}{h}\frac{Q^\ddagger[H^--HC_xN]}{Q[HC_xN]Q[H^-]}
  \label{rateeq}
\end{equation}

Here $k_B$ is the Boltzmann constant and $h$ is the Plank's constant. The PFs of the reaction complex are variationaly minimized along all the optimized geometries of the MEP. After the variational minimization we therefore obtain at each temperature the TS geometries at the lowest possible energy along the MEP paths and within the VTST model : see \cite{Fernandez-Ramos06}.

It is interesting to note  at this point that the rate coefficient as obtained from (\ref{rateeq}) is mainly controlled, at the lowest temperatures, by the behaviour of the vibrational part of the PF for that specific  TS geometry since all other degrees of freedom are now "frozen" to their lowest values. In turn, such low-temperature PFs relate directly to the zero-point energy (ZPE) values of the modes under consideration since $Q_{vib}$ ( at low T) can be written as  $\propto e^{-E_{ZPE}/KT}$. Thus the overall rates at low-T would be linked with the features of the corresponding ZPE energy values within the PFs:

\begin{equation}
  K(lowT)\propto T^2exp^{\frac{-E^{TS}_{ZPE}+E^{Reacts}_{ZPE}}{k_BT}}
  \propto T^2exp^{\frac{-\Delta_{ZPE}}{k_BT}}
  \label{kloweq}
 \end{equation}

in the present reactions we have that the  $E^{reacts}_{ZPE}=E^{HC_xN}_{ZPE}$ since $H^-$ does not contribute to it.
The quantity $\Delta_{ZPE}$  thus defines the energy differences between ZPEs of the reactants and those of the TS complex.It therefore controls the slope of the reaction rate K(lowT) which is defined by the above equation: a large value of that difference causes a greater decrease of K(T) as T decreases. The $\Delta_{ZPE}$ values of our present MEPs depend on: (i) the $H--C$ vibrations at the TS geometries and (ii) the  intramolecular vibrational frequencies within the remaining chain of atoms in the reactants. Since we found that the TS has a linear geometry for x=1, but it has increasingly more bent structures for x increasing from 3, to 5, to  7 (see Table \ref{tab6} and \ref{tab7} below), then the $\Delta_{ZPE}$ is the largest  for the HCN member of the series while it becomes increasingly  smaller for  increasingly larger members of the present series.
We report in detail  in Tables \ref{tab6} and \ref{tab7} all the geometric parameters for the TS structures we have found for our VTST calculations. We should note here that the calculations for x=1 and x=3 had been carried out already in our earlier work on the present mechanism \citep{satta15}, where we employed a smaller basis set expansion for the RPES calculations and carried out the VTST calculations over a less dense grid of points. The present, improved, results remain however within less than 5\% of the earlier ones and exhibit exactly the same temperature dependence.

  \begin{table}[H]
  \caption{Computed structures of the TS over a broad range of temperature for the smaller members of the present cyanopolyyne chains. Distances in \AA, angle in degrees}
\begin{centering}
  \begin{tabular}{|c|c|c|}
    \hline
    \multicolumn{3}{|c|}{HCN}\tabularnewline
\hline 
T & HH & HC\tabularnewline
\hline 
\hline 
10-340 & 1.721 & 1.055\tabularnewline
\hline 
345-475 & 1.712 & 1.059\tabularnewline
\hline 
480-500 & 1.702 & 1.062\tabularnewline
\hline 
\end{tabular}

\begin{tabular}{|c|c|c|c|c|c|}
  \hline
    \multicolumn{6}{|c|}{$HC_3N$}\tabularnewline
\hline 
T & HH & HC$_{1}$ & HHC$_{1}$ & HC$_{1}$C$_{2}$ & C$_{1}$C$_{2}$C$_{3}$\tabularnewline
\hline 
\hline 
10-255 & 1.713 & 1.030 & 158.12 & 166.31 & 179.04\tabularnewline
\hline 
260-350 & 1.702 & 1.034 & 158.38 & 166.00 & 179.01\tabularnewline
\hline 
355-440 & 1.692 & 1.037 & 158.64 & 165.69 & 178.98\tabularnewline
\hline 
445-500 & 1.681 & 1.041 & 158.88 & 165.37 & 178.95\tabularnewline
\hline 
\end{tabular}

\par\end{centering}
\label{tab6}
\end{table}

 \begin{table}[H]
  \caption{Computed structures of the TS over a broad range of temperature for the larger members of the present cyanopolyyne chains. Distances in \AA, angle in degrees}
\begin{centering}

\begin{tabular}{|c|c|c|c|c|c|c|}
  \hline
    \multicolumn{7}{|c|}{$HC_5N$}\tabularnewline
\hline 
T & HH & HC$_{1}$ & HHC$_{1}$ & HC$_{1}$C$_{2}$ & C$_{1}$C$_{2}$C$_{3}$ & C$_{2}$C$_{3}$C$_{4}$\tabularnewline
\hline 
\hline 
10-70 & 1.537 & 1.056 & 151.65 & 156.27 & 177.22 & 178.90\tabularnewline
\hline 
75-130 & 1.524 & 1.060 & 152.10 & 155.74 & 177.17 & 178.90\tabularnewline
\hline 
135-190 & 1.510 & 1.064 & 152.53 & 155.19 & 177.12 & 178.90\tabularnewline
\hline 
195-240 & 1.497 & 1.067 & 152.95 & 154.64 & 177.07 & 178.91\tabularnewline
\hline 
245-295 & 1.483 & 1.071 & 153.36 & 154.09 & 177.03 & 178.91\tabularnewline
\hline 
300-350 & 1.470 & 1.075 & 153.75 & 153.53 & 176.99 & 178.91\tabularnewline
\hline 
355-410 & 1.456 & 1.079 & 154.12 & 152.97 & 176.95 & 178.91\tabularnewline
\hline 
415-480 & 1.442 & 1.082 & 154.49 & 152.41 & 176.91 & 178.91\tabularnewline
\hline 
485-500 & 1.429 & 1.086 & 154.84 & 151.84 & 176.88 & 178.91\tabularnewline
\hline 
\end{tabular}

  \begin{tabular}{|c|c|c|c|c|c|c|}
\hline
    \multicolumn{7}{|c|}{$HC_7N$}\tabularnewline
    \hline
T & HH & HC$_{1}$ & HHC$_{1}$ & HC$_{1}$C$_{2}$ & C$_{1}$C$_{2}$C$_{3}$ & C$_{2}$C$_{3}$C$_{4}$\tabularnewline
\hline 
10-35 & 1.476 & 1.062 & 148.39 & 153.34 & 176.52 & 178.64\tabularnewline
\hline 
40-75 & 1.462 & 1.066 & 148.96 & 152.69 & 176.46 & 178.64\tabularnewline
\hline 
80-115 & 1.447 & 1.070 & 149.51 & 151.04 & 176.40 & 178.65\tabularnewline
\hline 
120-165 & 1.433 & 1.074 & 150.04 & 151.40 & 176.35 & 178.65\tabularnewline
\hline 
170-210 & 1.418 & 1.077 & 150.55 & 150.76 & 176.30 & 178.65\tabularnewline
\hline 
215-270 & 1.404 & 1.081 & 151.03 & 150.12 & 176.26 & 178.66\tabularnewline
\hline 
275-315 & 1.389 & 1.085 & 151.50 & 149.47 & 176.22 & 178.66\tabularnewline
\hline 
320-365 & 1.375 & 1.089 & 151.95 & 148.84 & 176.19 & 178.66\tabularnewline
\hline 
370-420 & 1.361 & 1.092 & 152.37 & 148.21 & 176.17 & 178.67\tabularnewline
\hline 
425-465 & 1.347 & 1.096 & 152.79 & 147.58 & 176.14 & 178.67\tabularnewline
\hline 
470-500 & 1.333 & 1.100 & 153.19 & 146.96 & 176.13 & 178.67\tabularnewline
\hline 
\end{tabular}
\par\end{centering}
\label{tab7}
\end{table}

The data of these Tables show, from the top to the bottom panels, the x=1 to x=7 cases. The first two columns report distances of the atoms in the complex: the $H^--H$ and $H-C_1$, while the next columns give angles along the bent structures of the longer members of the series. Only the $H^--HCN$  reaction shows a linear TS structure.

Since, within the RRKM approach which we have used in the present modeling, the structure of the TS is crucially linked to the expression for the resulting rate coefficient via eq.s (\ref{rateeq}) and (\ref{kloweq}), it is interesting to note the following from a perusal of the data reported in the above Tables:

\begin{enumerate}[label=(\roman*)]
\item The structures of the TSs change little over a very broad range of temperatures, indicating that the variational determination of the TS is correctly obtained: along an exothermic, barrierless path the structure of the final complex should change little from its minimum structure;
\item as the length of the chain increases we see more marked departures from linearity along what we have defined as the "translational" coordinates which are active along the MEP and which involve the X-C-H-H bonds, while the C-atoms further away seem to maintain a nearly linear structure.
  It means that only the nearest triple bond is affected by the polarization changes during  the ET process, while those further away along the chain remain largely unchanged, as expected;
  \item The modification of that triple bond, on the occurrence of excess charge initial transfer, which reaches its final, main localization at the terminal N atom of the product anion, also affects the $\Delta_{ZPE}$ energy gaps discussed before, a feature which is crucial for the low-T behaviour of the rate coefficients (see eq.s \ref{kloweq}): it follows then that the $\Delta_{ZPE}$ value increases again from x=3 to x=5 and x=7. This feature will be further discussed below when analysing the low-temperature  behaviour of the present rates.
\end{enumerate}

We have further fitted all the computed rates following the standard formula suggested earlier for such systems, see: \cite{gian16, satta15, prasad80} and we therefore report the actual fitting parameters in Table \ref{tab8} for the higher-T range (upper panel) and the low-T range (lower panel). The actual fitting formula is also reported at the bottom of that Table.

The data for the x=1 member of the series will be discussed later on comparison with the HNC isomeric  variant.

\begin{table}[H]
\begin{centering}
\begin{tabular}{|c|c|c|c|c|}
\hline 
T (K) & n & $\alpha$ (cm$^{3}$/s) & $\beta$ & $\gamma$ (K)\tabularnewline
\hline 
\hline 
\multirow{4}*{100-500} & 1 & 9.0368e-10 & 1.7361 & 284.429\tabularnewline
\cline{2-5} 
 & 3 & 2.3052e-9 & 1.7589e-1 & 153.699\tabularnewline
\cline{2-5} 
 & 5 & 2.7539e-9 & -3.7352e-2 & 188.310\tabularnewline
\cline{2-5} 
 & 7 & 3.1121e-9 & -3.9163e-1 & 391.083\tabularnewline
\hline 
\multirow{4}*{10-100} & 1 & 1.8152e-9 & 6.4397e-1 & 479.034\tabularnewline
\cline{2-5} 
 & 3 & 2.4787e-9 & 4.9065e-2 & 173.248\tabularnewline
\cline{2-5} 
 & 5 & 2.7820e-9 & -4.9593e-2 & 190.971\tabularnewline
\cline{2-5} 
 & 7 & 2.8484e-9 & -2.4470e-1 & 367.335\tabularnewline
\hline
\multicolumn{5}{|l|}{$K(T)=\alpha\left(\frac{T}{300}\right)^{\beta}e^{-\frac{\gamma}{T}}$}\tabularnewline
\hline
\end{tabular}
\par\end{centering}
\caption{Fitting parameters for the rate coefficients computed in the present work,  given for the high-T range (upper panel) and the low-T range (lower panel). The actual fitting formula is given at the bottom of the Table.}
\label{tab8}
\end{table}

\begin{figure}[H]
\includegraphics [height=15cm] {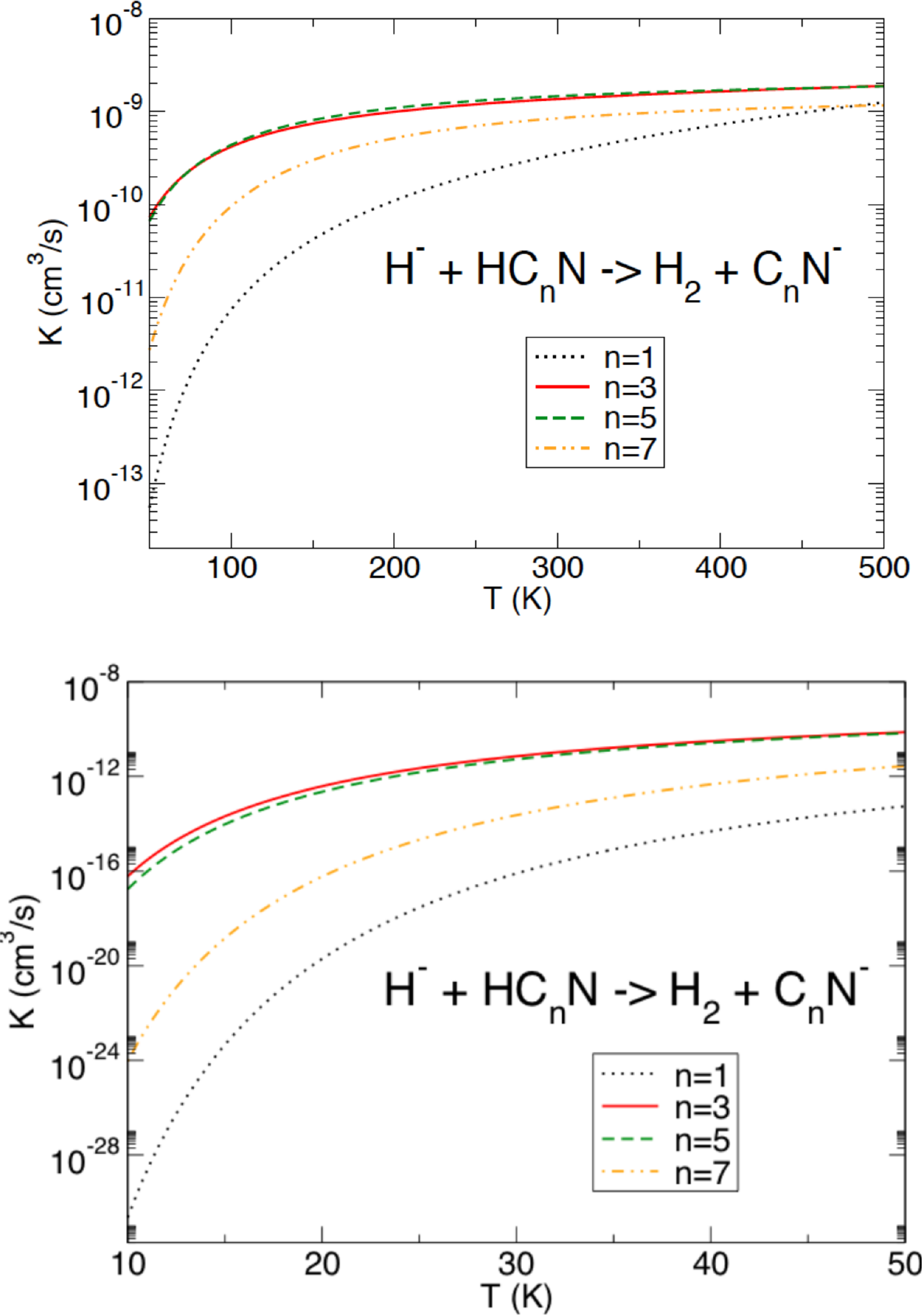}
\caption{Computed temperature dependence of the reaction rates of the present study from $CN^-$ to $C_7N^-$ for the odd values of the $x$ index. The lower panel reports the low-T behaviour, while the upper panel gives the higher-T behaviour of the rates of formation.}
\label{fig5}
  \end{figure}

The temperature behaviour of the computed rates is reported for all present systems  by the two panels of Figure \ref{fig5}

  The following comments could be made by observing the rate behaviour in that figure:

  \begin{enumerate}[label=(\roman*)]
\item as expected, the formation of $CN^-$ remains the least efficient chemical process in reaction with $H^-$. All other members of the series exhibit larger reaction rates at all temperatures. As discussed earlier, this feature has to do with the instability of the anion of the molecular reaction partner and therefore with the importance of initiating  the ET step only at  large bond distances for the H--CN fragment of the reacting cyanopolyyne, where the final hydrogen molecule has not yet been formed.  At low temperature that step may not occur while at the higher temperature its increased occurrence is still limited by the fewer degrees of freedom of the PF of its transition complex;
\item The largest rates are seen to occur for the formation of $C_3N^-$ and, close to it in size, by the formation of $C_5N^-$. These are the two species exhibiting the largest rates of formation from the present calculations. Since both systems were shown to have TS complexes not far from the more efficient linear structures, it stands to reason that both molecular anions are formed with larger rates with respect to those of the x=1 system, where the reduced number of degrees of freedom involved in its PFs limits the size of the corresponding rates;
\item The largest member of the series, the $C_7N^-$ shows a smaller rate of formation at all T,  smaller than the x=3 and 5 members but still larger than the x=1 member. This difference has to do with the fact, already discussed earlier, that the TS variationally  optimized structures are now far from the linear configurations and therefore the bending of the longer chain attached to the active coordinates reduces the numerator in (\ref{rateeq}), which in turn lowers the value of the associated reaction rate. It is further worth noting here that, with the exclusion for the moment of the x=1 member,  the x=3 and 5 anions are seen in our calculations to be associated with largest rates of reaction. They  are  two members of the present series of molecules for which the anions have been repeatedly detected in the ISM, while $C_7N^-$  larger member, so far, has not yet been detected by observational studies (e.g. see: ref \cite{mill17};
  \item One should also note that the Langevin rates listed in the KIDA database of \cite{wake12} for $CN^-$ are about $3.8\cdot10^{-9}$ cm$^3$ mol$^{-1}$ s$^{-1}$ and independent of temperature. They are obviously always overestimating all present rates and at all temperatures.
\end{enumerate}

Another interesting set of comparison between the values of the rates of formation, at a four selected temperatures indicative of different regions of the ISM are reported in Table \ref{tab9} for the four members of the cyanopolyyne series of this work.

\begin{table}[H]
\begin{centering}
\begin{tabular}{|c|c|c|c|c|}
\hline 
n & 10 K & 30 K & 50 K & 100 K\tabularnewline
\hline 
\hline 
1 & 2.29(-31) & 7.85(-17) & 5.53(-14) & 7.40(-12)\tabularnewline
\hline 
3 & 5.96(-17) & 7.10(-12) & 7.34(-11) & 4.19(-10)\tabularnewline
\hline 
5 & 1.71(-17) & 5.29(-12) & 6.67(-11) & 4.38(-10)\tabularnewline
\hline 
7 & 8.58(-25) & 2.26(-14) & 2.75(-12) & 9.57(-11)\tabularnewline
\hline 
\end{tabular}
\par\end{centering}
\caption{Computed rate values, at four different temperatures, for the present series of cyanopolyynes  (units of  cm$^3$ mol$^{-1}$ s$^{-1}$)}
\label{tab9}
\end{table}

One sees that at the lowest shown  temperature of 10 K the formation of $CN^-$ and $C_7N^-$ are really negligible, while the formation of $C_3N^-$ and $C_5N^-$ is still small but many orders of magnitude larger. The effects are more marked for the region between 30 and 100 K, indicatively corresponding to the averaged temperatures of Dark Molecular Clouds and CSE environments, see: \cite{mill17}. In these intervals we see that the rates of formation become much larger for all four systems, albeit still being largest for $C_3N^-$ and $C_5N^-$.  Such data indicate that the possibility of forming all anions by chemical routes becomes significant for the latter molecules and in clear competition with REA processes which have been shown as  having  much smaller rates of formation associated with them. One should also be reminded that for temperatures around 50 K the chemical rates of formation of the x=1 member of the present series are already orders of magnitude larger than its calculated REA rates around 100 K : \cite {kames16}. It therefore follows that even for the smallest member of the present series of anions the chemical paths reported here is much more significant than the REA mechanism for explaining the formation of its anion .

\begin{figure}[H]
\includegraphics[height=15cm]   {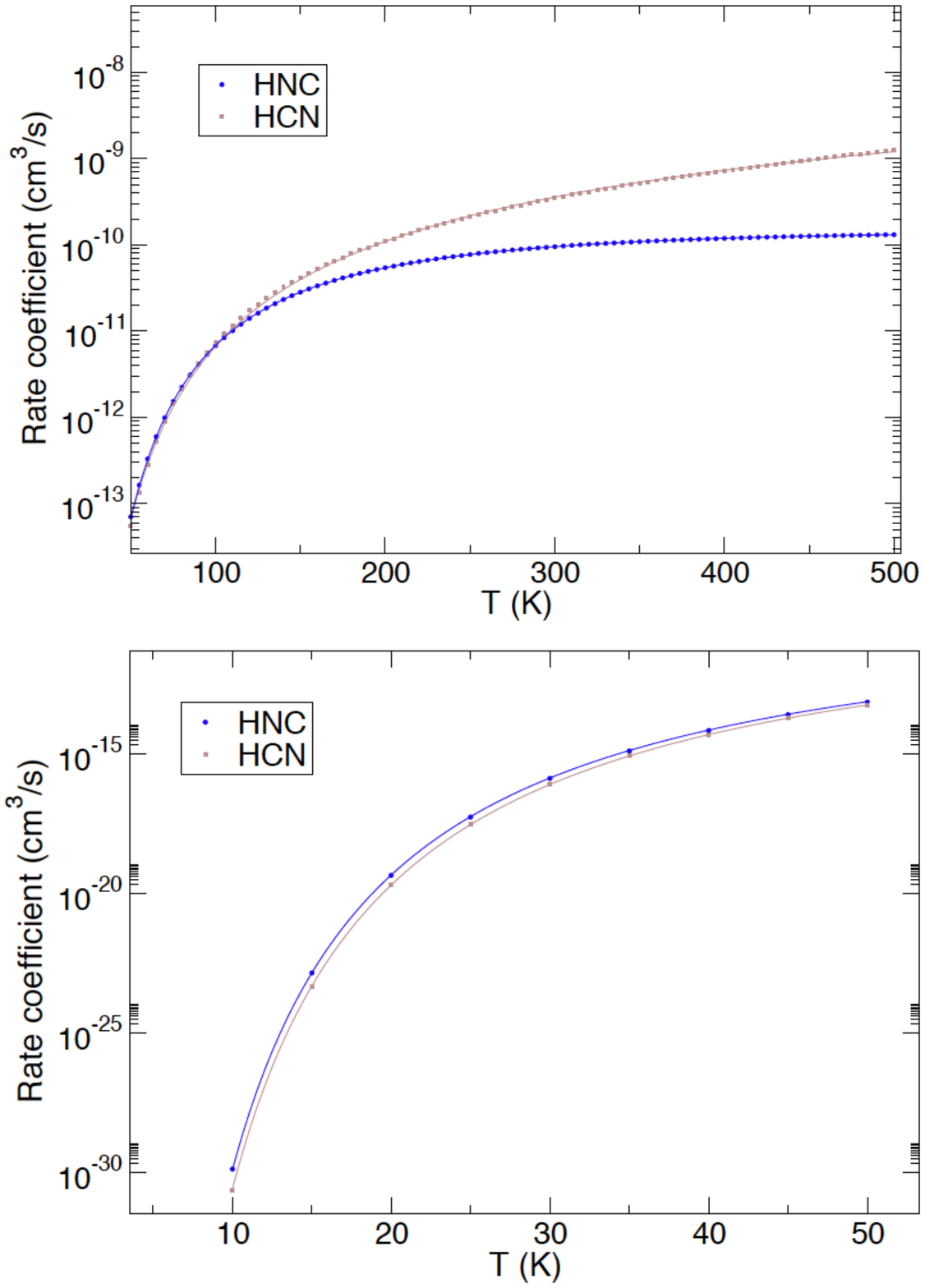}
\caption{Computed formation rates of $CN^-$ form both isomeric forms of the initial reagent: HCN  and HNC. See main text for further details. Lower panel: low-T range; upper panel: high-T range}
\label{fig6}
 \end{figure}

Another interesting comparison between some of our data is shown in the panels of Figure \ref{fig6}, where the rates of formation of $CN^-$ are shown for both the isomer partners of $H^-$: HCN and HNC.

 It is interesting to note that both  reagents, as expected, behave very similarly as a function of temperature. The low-T formation rate are very close to each other, although the $CN^-$ formation rate from HNC remains always slightly larger.
 As the temperature moves above 150 K we see that the differences in size increase and the trend is inverted.
 We see in the upper panel of fig. \ref{fig6} that the HNC  reagent yields formation rates which are increasingly smaller, reaching nearly one order of magnitude smaller around 500 K. On the whole, however, the rates of formation at 200 K  and above remain of the order of $10^{-10}$ cm$^3$ mol$^{-1}$ s$^{-1}$, i.e. substantially larger than the corresponding rates of REA formation by electron attachment \cite{herbst09, Kawaguchi95, kames16}. All reaction rates, however, remain smaller than the Langevin formation rates usually employed in astrochemical databases and which are, as mentioned earlier, of the order of $10^{-9}$ cm$^3$ mol$^{-1}$ s$^{-1}$ as given by  \cite{wake12}.

 We summarize in Table \ref{tab10} the fitting parameters for the two isomeric partners of  the smallest member of the present series of cyanoderivatives, both leading to the formation of $CN^-$.
 
 \begin{table}[H]
\caption{Computed fitting parameters for the T-dependence of the formation rates of $CN^{-}$ from HNC (left) and HCN (right). The formula is the same as that of Tables \ref{tab6} and \ref{tab7}. All rates in units of cm$^{3}$ mol$^{-1}$ s$^{-1}$.}
\begin{tabular}{|c|c|c|c|c|}
\hline 
 & \multicolumn{2}{c|}{HNC} & \multicolumn{2}{c|}{HCN}\tabularnewline
\hline 
\hline 
T(K) & 10-50 & 50-500 & 10-50 & 50-500\tabularnewline
\hline 
$\alpha(cm^{3}/s)$ & 4.9394$\cdot10^{-10}$ & 5.2509$\cdot10^{-10}$ & 6.9512$\cdot10^{-10}$ & 1.1986$\cdot10^{-9}$\tabularnewline
\hline 
$\beta$ & -0.6029 & -0.7082 & -0.3966 & 1.4335\tabularnewline
\hline 
$\gamma(K)$ & 495.3 & 511.6 & 507.5 & 362.1\tabularnewline
\hline 
\end{tabular}
\label{tab10}
 \end{table}

 It is also interesting to comment on the possible reasons for the difference in the temperature dependence of the two rates.
 We already know that the HC bond is less strong than the HN bond in its isomeric variant, as shown by our data discussed in Section II.
 We have already discussed before that, at low-T,  the present chemical rates are controlled by the differences in ZPE values between the two TS (linear) for both molecules. Since the one for HCN is larger than for HNC, then the low-T formation rates given by our VTS theory slightly favour the formation from HCN than that from HNC. As the temperature increases, however, the contribution from the CN frequencies in HCN vs HNC are higher in the former with respect to the latter. Furthermore, the opposite occurs for the HC frequency in the TS with respect to the HN frequency: it is about 3311 cm$^{-1}$ in the former and 3643 cm$^{-1}$ in the latter. Thus, as the temperature increases the stronger HN bond within the TS complex slows down the rate increase for HNC vs those for HCN.
 On the whole, however, both molecules can contribute with similar rates to $CN^-$ formation, thereby making its probability of occurrence even larger than via the REA route (about $10^{-17}$ cm$^{3}$ mol$^{-1}$ s$^{-1}$ around 100 K: \cite{herbst09, kames16})

 \section{Present Conclusions}
 In the present work  we have analysed in detail the chemical formation of $C_xN^-$ anionic species from a  reaction with $H^-$ of the corresponding cyanopolyyne species: $HC_xN$, with x from 1 to 7. The aim has been to evaluate, for temperatures of significance under ISM conditions, the relative probabilities of forming the anions and to show that this specific  chemical route is likely to be  more significant than  the more popular electron attachment processes followed by radiative emission which has been considered for a long time as the chief source of those anions, i.e. the REA formation mechanism already discussed  many times in the literature \cite{herbst09, mackay77, care14, carelli13, kames16}.

 We have carried out structural calculations for the reagents with x=1, 3, 5, 7 and further employed  accurate ab initio findings to generate the RPES along the collinear path. We have further calculated the nonlinear MEP to product formation. The behaviour of all the above structural quantities along the series of cyanopolyynes has been discussed in detail in sections II and III.

 In section IV we have employed the computed MEP to obtain the anion formation rates of reaction using the VTS theory and examining the effects of nonlinear TS for the longer chains of the series. We thus found that while $CN^-$ is formed via a linear MEP to final RRKM TS that in turns decays into products,   the longer chains with x=3, 5 and 7 show an increasing importance of non linear TS structures for the formation of the complexes, a feature that finally produces a strongly bent transition complex for the x=7 member of the series, and  corresponding rates of  formation which are lower than those of the two preceeding,  smaller terms of the same series.

 The calculations of the reaction rates also show a distinct behaviour as a function of temperature: (i) a very rapid increase of several orders of magnitude when going from a few K to about 100 K, and (ii) a smoother increase with temperature when T increases up to 500-1000 K. This behaviour was shown here to be linked to the importance of the ZPE differences between reactants and their TS structures . We found in fact them  to be smallest for the $C_3N^-$ formation reaction, so that  the latter rates  show the largest values in the low-T regimes.

 All reactions were  found to be strongly exothermic, without a barrier from reactants to products and forming a TS complex along the MEP. It is along that path  that the reactants undergo $H_2$ formation in its lowest vibrational state, following an ET process that begins to take place  when forming an initial complex with a highly stretched   $H-H$ reagent and finally evolves to the anionic  $C_xN^-$. product The computed rate values reported by Table \ref{tab9} show that in going from 10 K to 100 K  they change differently  for each member of the series : the least reactive HCN changes nearly twenty orders of magnitude, while the more efficient formations of $C_3N^-$ and $C_5N^-$ vary by about seven orders of magnitude: the changes with temperature therefore turn out to be very dramatic, for reasons explained in the previous Section.

 In the range of temperatures likely to be present  in CSE environments, where those anions have been detected (e.g. see the Review by \cite{mill17}) we therefore see that the present chemical process for $C_3N^-$ is several orders of magnitude larger than the REA rate of formation \cite{mackay77, kames16}.

 Furthermore, we also see that the size of the Langevin rates of anion formation, usually employed in modelling studies: \cite{wake12}, are invariably larger than any of our computed rates. It indicates the low reliability of using such estimates for the present  ionic reactions.

 Another interesting result from the present calculations is the analysis of the rates of formation for the HCN/HNC isomers. Their relative abundances  in the ISM have been discovered to be very different from the one on earth, thus suggesting that both species could contribute equally to the production of $CN^-$ anions. In fact, our calculations for both species, presented in Figure \ref{fig6}, indicate their rates to exhibit some differences as a function of temperature, but to be essentially of the same order of magnitude around 100-200 K.
 Given the fact that the REA path of formation was found to be several orders of magnitude smaller over the same range of temperature \cite{kames16}, our findings suggest again that the present chemical route to $CN^-$ formation is a more efficient mechanism for forming that anion in the CSE environments or dark molecular clouds.

 The present calculations further indicate that the x=7 case produces RRKM rates of formation which are smaller than those of the shorter chains with x=3 and 5, this being more so in the lower-T range below 50 K. The specific structural reason provided by our calculations indicate that the TS complex along the MEP for this reaction favours strongly bent structures which therefore increase the importance of bending vibrations within the PF of the complex. This results into a reduction of the rate values obtained within the VTST model.

Such finding is in line with the fact that, thus far,the  $C_7N^-$ anionic molecule has not been observed although $HC_7N$ has been observed: \cite{mill17}. on the other hand,  both $C_3N^-$ and $C_5N^-$ have been detected by \cite{gupta07, herbst81}.

 The following conclusions could therefore be drawn from the results of the present calculations:
 
 \begin{enumerate}[label=(\roman*)]
 \item The chemical paths to anionic derivatives from initially neutral  cyanopolyyne  in reaction with  $H^-$   are shown to be more effective than the REA paths (whenever available) at the ISM temperature of interest;
 \item The $C_3N^-$ and  $C_5N^-$ formation exhibit the largest chemical rates in the range of 50 K to 100 K and therefore the present reactions are shown to be very effective mechanisms for their formation;
 \item The formation of $CN^-$ can occur via two different reagents, HCN and HNC, which were shown to have very similar rates. Albeit being the smallest found by our calculations, their combined occurrence can be markedly larger than the existing REA rates indicated in the current literature \cite{kames16}, thereby making this chemical reaction an important path to their formation;
   \item  the $C_7N^-$ formation rates  turn out to be smaller than the preciding shorter members of the series, thus suggesting this chemical "inefficiency" as a possible cause for the lack of its observational detection as discussed  in \cite{mill17}.   
   \end{enumerate}

The present chemical study therefore confirms from ab initio calculations that the analysis of ion-molecule reactions involving $H^-$ should be experimentally attempted  with cyanopolyynes  to search for chemical routes to the  formation of anionic  (C,N)-containing linear chains as those observed already in different ISM  environments \cite{mill17}.
 
\section*{Acknowledgements}

F.A.G. and R.W. thank the  Austrian Science Fund (FWF) for supporting the present research through the Project P27047-N20. 

\begin{small}

\end{small}

\end{document}